\documentclass[3p,times]{elsarticle}

\usepackage{ecrc}

\usepackage{epstopdf}

\volume{00}

\firstpage{1}

\journalname{Nuclear Physics A}

\runauth{Roberta Arnaldi for the ALICE Collaboration}

\jid{nupha}

\jnltitlelogo{Nuclear Physics A}

\usepackage{graphicx}
\usepackage{amsmath,amssymb}

\newcommand{\jpsi}{\rm J/$\psi$}
\newcommand{\psip}{$\psi(\rm 2S)$}

\newcommand{\pt}{$p_{{\mathrm T}}$}

\newcommand{\RpPb}{$R_{\mathrm{pPb}}$}

\newcommand{\QpPb}{$Q_{\mathrm{pPb}}$}

\newcommand{\sqrts}{$\sqrt{s_{\rm NN}} = 5.02$ TeV}

\begin{document}


\begin{frontmatter}



\title{Inclusive $\psi$(2S) production in \mbox{p-Pb} collisions with ALICE}

\author{Roberta Arnaldi (for the ALICE\fnref{col1} Collaboration)}
\fntext[col1] {A list of members of the ALICE Collaboration and acknowledgements 
can be found at the end of this issue.}
\address{INFN Torino, via P. Giuria 1, I-10125 Torino (Italy)}




\begin{abstract}
The ALICE Collaboration has studied the inclusive \psip\ production in 
\mbox{p-Pb} collisions at \sqrts\ at the CERN LHC. 
Measurements are performed, in the $\mu^{+}\mu^{-}$ decay channel, in a forward (2.03$<y_{\rm cms}<$3.53) and in a backward (-4.46$<y_{\rm cms}<$-2.96) centre of mass rapidity ranges, as a function of transverse momentum or event activity. 
The \psip\ production is compared to the \jpsi\ one through the double ratio $[\sigma_{\psi(\rm 2S)}/\sigma_{\rm J/\psi}]_{\rm pPb}/[\sigma_{\psi(\rm 2S)}/\sigma_{\rm J/\psi}]_{\rm pp}$ between the cross 
sections evaluated in \mbox{p-Pb} and \mbox{pp} collisions and by calculating the \jpsi\ and 
\psip\ nuclear modification factors. 
Results indicate that the \psip\ production is suppressed with respect to 
\mbox{pp} and, in particular in the backward rapidity region, the suppression is stronger 
than the \jpsi\ one. This unexpected difference between the \jpsi\ and \psip\ behaviour, not accounted for by 
theoretical models based on nuclear parton shadowing and coherent energy loss mechanisms, indicates 
the presence of sizeable final state effects on the \psip\ production.
\end{abstract}


\end{frontmatter}



\section{Introduction}
\label{intro}
The high density of colour charges, present in a plasma of quarks and gluons (QGP) created in heavy-ion 
collisions, is expected to screen the binding between heavy (charm or beauty) $Q$ and $\overline Q$ quarks, 
suppressing the  production of quarkonium states with respect to the production in \mbox{pp}~\cite{Mat86}. 
In this scenario, the different binding energies of the various $Q\overline Q$ states should lead to a sequential suppression pattern, where, in case of charmonium resonances, the more loosely bound \psip\ melts at lower temperatures with respect to the more tightly bound \jpsi. 
The NA50 experiment has indeed observed a stronger \psip\ suppression relative to the \jpsi\ one in \mbox{Pb-Pb} collisions at $\sqrt{s_{\rm NN}}=17$ GeV~\cite{Ale07}. 
However, a strong modification of the \psip\ yields was measured also in \mbox{p-A} collisions~\cite{Ale06, Lei00,Abt07}, where no QGP formation is expected. In particular, a stronger \psip\ suppression relative to the \jpsi\ was observed at central rapidity ($y$), while at forward-$y$ the two resonances followed a similar trend.
This behaviour is interpreted in terms of cold nuclear matter (CNM) effects, such as nuclear parton shadowing, energy loss and $c\overline c$ break-up in interactions with nucleons. 
While the first two mechanisms are not expected to significantly depend on the charmonium state, 
the $c\overline c$ break-up probability is sensitive to the size of the object crossing the medium.
More in details, while the coloured $c\overline c$ pair produced by gluon fusion evolves first into a colour neutral object and eventually into a fully formed resonance~\cite{Arl00}, its size grows, increasing the break-up 
probability. If the charmonium formation time ($\tau_{\rm f}$) is smaller than the nucleus crossing time ($\tau_{\rm c}$), the object experiencing the nuclear medium is a fully formed resonance which can be suppressed according to its binding energy.
On the contrary, if the resonance forms outside the nucleus, i.e. if a $c\overline c$ pair not yet evolved into a 
\jpsi\ or \psip\ crosses the medium, a similar behaviour is expected, independently of the final charmonium state. 
As the collision energy increases, the time spent by the $c\overline c$ pair inside the nucleus
decreases due to its large Lorentz-$\gamma$ factor. The $c\overline c$ break-up contribution
becomes negligible and a similar \jpsi\ and \psip\ suppression is envisaged. In contrast to these expectations, 
the PHENIX Collaboration has observed a stronger \psip\ suppression relative to the \jpsi\ in \mbox{d-Au} collisions at $\sqrt{s_{\rm NN}} = 200$ GeV~\cite{Ada13}. 
ALICE has now addressed the study of the \psip\ production in \mbox{p-A} 
collisions at LHC energies to shed some light on this observation and to clarify the role of CNM effects.

\section{Analysis and physics results}
\label{analysis}
The ALICE Collaboration has studied both \jpsi\ and  \psip\ production in \mbox{p-Pb} collisions at
\sqrts~\cite{Abe04,Abe04pA}. Charmonium resonances are measured, through their dimuon decay channel, in the
Muon Spectrometer, covering the pseudorapidity range $-4<\eta_{\rm lab} <-2.5$. 
The two innermost layers of the Inner Tracking System provide the vertex identification, 
and two scintillator hodoscopes (VZERO) are used for triggering purposes.  A set of Zero Degree 
Calorimeters (ZDC) helps to remove de-bunched collisions and to determine the event activity. 
Details on the ALICE experimental apparatus can be found in~\cite{Aam08}. 
Data have been collected under two different configurations, inverting the direction of
the p and Pb beams. In this way both forward ($2.03<y_{\rm cms}<3.53$) and backward ($-4.46<y_{\rm cms}<-2.96$) 
centre of mass rapidities could be accessed, with the positive $y$ defined in the direction of the proton beam. 
The difference in the covered $y$ ranges reflects the shift of the centre of mass of the nucleon-nucleon collisions ($\Delta y = 0.465$) with respect to the laboratory frame, induced by the different energies per nucleon of the colliding beams.
The resonance yields are extracted by fitting the dimuon invariant mass distributions with a superposition of signals and background shapes. For the signal pseudo-Gaussian or Crystal Ball functions, with asymmetric tails on both sides of the resonance peak, are used, while for the background a Gaussian with a mass-dependent width or polynomial $\times$ exponential functions are adopted.
The \psip\ yield and its statistical uncertainty is obtained as the average of the results of the fits performed combining the various signal and background shapes, while the systematic uncertainty is given by the RMS of the distribution of the fit results. 
A total number of $N_{\rm \psi(2S)}$=1069$\pm$130(stat)$\pm$102(syst) is obtained at forward-$y$, while at backward-$y$ the corresponding figure is $N_{\rm \psi(2S)}$=697$\pm$111(stat)$\pm$65(syst). 
The \psip\ yields are then divided by the acceptance $\times$ efficiency ($A\times\epsilon$) evaluated using Monte-Carlo simulations. 
The \psip\ $A\times\epsilon$ values, averaged over \pt\ and $y$, are $(27.0 \pm 1.4)$\%  and $(18.4 \pm 1.3)$\%  at forward and backward $y$ respectively, where the quoted uncertainties are systematic.
Details on the signal extraction, $A\times\epsilon$ correction and systematic uncertainties can be found in ~\cite{Abe04,Abe04pA}.
\begin{figure}[htbp]
\begin{center}
\includegraphics*[width=0.45\textwidth]{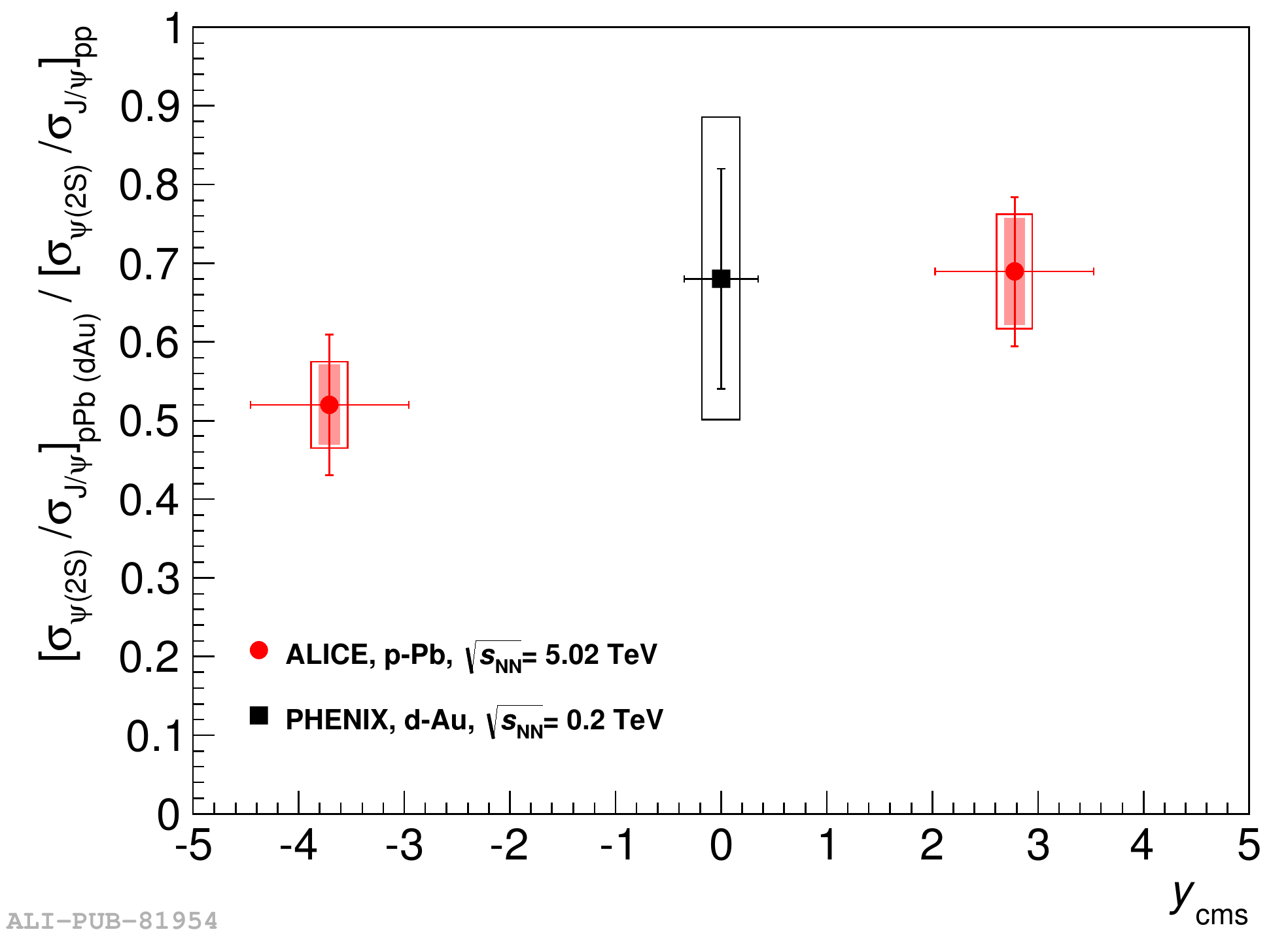}
\includegraphics*[width=0.45\textwidth]{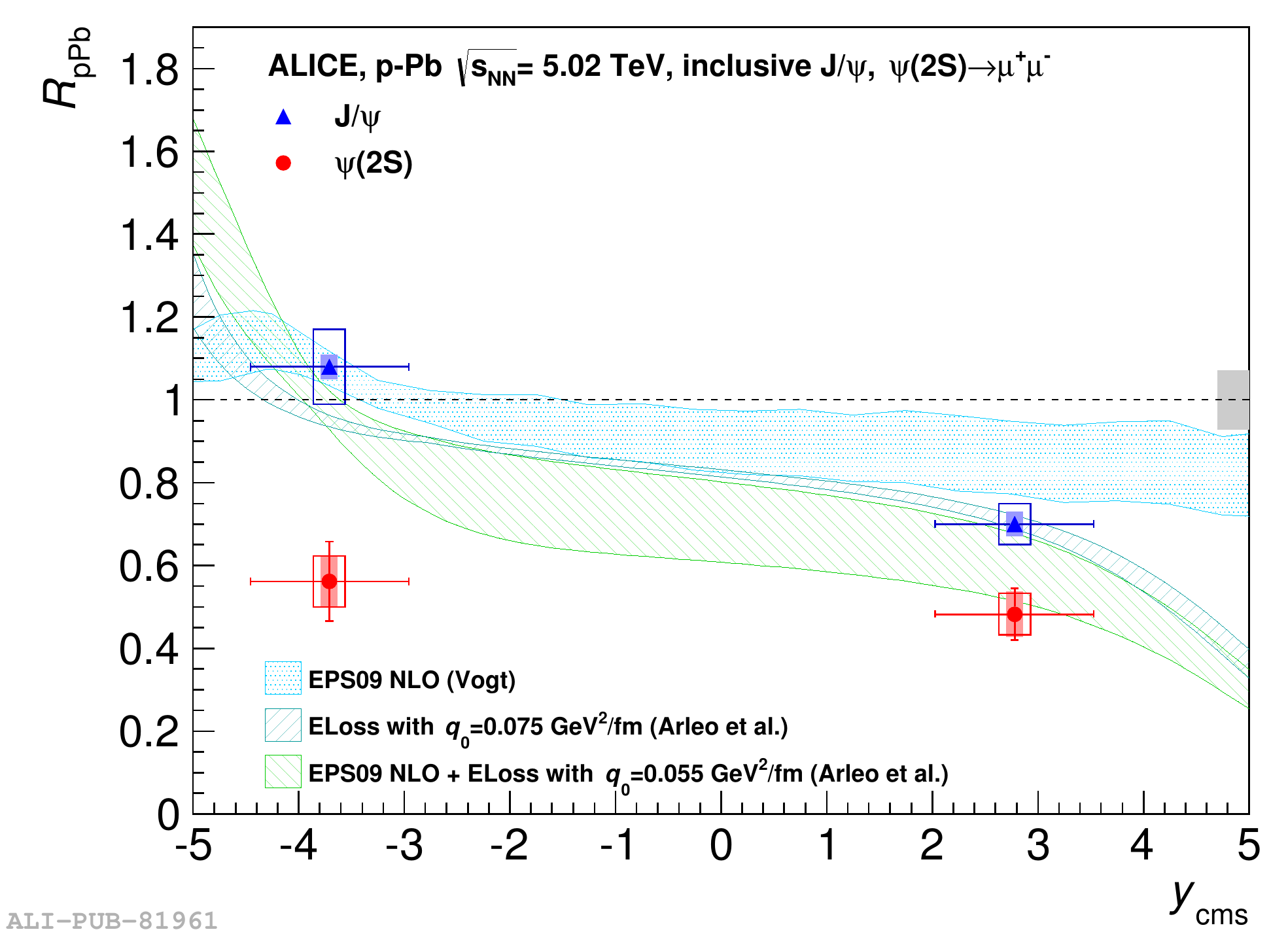}
\caption{
Left: Double ratios $[\sigma_{\psi(\rm 2S)}/\sigma_{\rm J/\psi}]_{\rm pPb}/[\sigma_{\psi(\rm 2S)}/\sigma_{\rm J/\psi}]_{\rm pp}$ 
for \mbox{p-Pb}, as a function of $y$, compared to the corresponding PHENIX result~\cite{Ada13}. 
The horizontal bars show the width of the $y$ regions under study. For ALICE, the vertical error bars correspond to statistical uncertainties, the boxes (shaded areas) to uncorrelated (correlated) systematic uncertainties. 
Right: \psip\ and \jpsi\ \RpPb\ versus $y$ compared to theoretical models. Vertical error bars correspond to statistical uncertainties, boxes to uncorrelated systematic uncertainties, and shaded areas to partially correlated uncertainties. 
The filled box on the right, centered on \RpPb=1, corresponds to fully correlated uncertainties between \jpsi\ and \psip.
}
\label{fig:fig1}
\end{center}
\end{figure}
The \psip\ production cross section times the branching ratio is computed as $ {\rm B.R.}\cdot\sigma^{\rm \psi(\rm 2S)}_{\rm pPb}=N^{\rm cor}_{\rm \psi(\rm 2S)}/N_{\rm MB}\times{\sigma^{\rm MB}_{\rm pPb}}$, where ${N^{\rm cor}_{\psi(\rm 2S)}}$ is the number of \psip\ corrected for $\rm{A}\times \epsilon$, $N_{\rm MB}$ is the number of minimum bias (MB) events and $\sigma^{\rm MB}_{\rm pPb}$ is the cross section for the occurrence of the MB condition~\cite{Abe04pA}. 
The \psip\ production cross section in \mbox{p-Pb} is compared to the \jpsi\ one and to the corresponding quantities in \mbox{pp} interactions in terms of double ratio 
$[\sigma_{\psi(\rm 2S)}/\sigma_{\rm J/\psi}]_{\rm pPb}/[\sigma_{\psi(\rm 2S)}/\sigma_{\rm J/\psi}]_{\rm pp}$. 
Since no \mbox{pp} data are available at \sqrts, the results obtained at $\sqrt{s_{\rm NN}} = 7$ TeV, in $2.5<y_{\rm cms}<4$, have been used~\cite{Lop14}.
A 8\% systematic uncertainty on the double ratio is included to take into account the different $\sqrt{s}$ 
and $y$ range~\cite{Abe04pA} where the $[\sigma_{\psi(\rm 2S)}/\sigma_{\rm J/\psi}]$ is measured.
Results obtained at forward and backward $y$ are shown in Fig.~\ref{fig:fig1}(left), where a strong decrease of the \psip\ yields, relative to the \jpsi, is observed in both $y$ ranges.
The \psip\ to \jpsi\ double ratio is also compared to the result obtained by PHENIX in \mbox{d-Au} collisions at $\sqrt{s_{\rm NN}} = 200$ GeV~\cite{Ada13}. 
ALICE results show that, compared to pp, the \psip\ is more suppressed than the J/$\psi$ to a 2.1$\sigma$ (3.5$\sigma$) level at forward (backward) $y$, while the PHENIX value shows a similar feature, at a 1.3$\sigma$ level. 

The charmonium suppression with respect to the corresponding \mbox{pp} yield is quantified through the nuclear modification factor \RpPb. The \psip\ \RpPb\ is obtained from the aforementioned double ratio and from the \jpsi\ 
\RpPb~\cite{Abe04pA}, as $R^{\psi(\rm 2S)}_{\rm pPb}=R^{{\rm J}/\psi}_{\rm pPb}/\big[(\sigma^{\psi(\rm 2S)}_{\rm pPb}/
\sigma^{{\rm J}/\psi}_{\rm pPb})\cdot(\sigma^{{\rm J}/\psi}_{\rm pp}/\sigma^{\psi(\rm 2S)}_{\rm pp})\big]$.
Results, shown in Fig.~\ref{fig:fig1}(right), indicate a stronger \psip\ suppression with respect to the \jpsi, reaching a factor 2 relative to \mbox{pp}. 
\RpPb\ are also compared to theoretical calculations based on nuclear shadowing~\cite{Alb13} or coherent energy loss, with or without shadowing contribution~\cite{Arl13}. 
Since the kinematic distributions of gluons producing the \jpsi\ or the \psip\ are rather similar and since the coherent energy loss does not depend on the final quantum numbers of the resonances, the same theoretical calculations hold for both \jpsi\ and \psip.
Theoretical models predict a $y$ dependence which is in fair agreement with the \jpsi\ data, but in 
strong contradiction with the \psip\ result, indicating that other mechanisms should be invoked to describe 
the \psip\ production in \mbox{p-Pb}.
The resonance break-up cross section in the medium depends on the binding energy of the charmonium state and could be, in principle, a natural explanation for the experimental observation of the different \jpsi\ and \psip\ suppression, as for low energy results.
However, this process can only play a role if the charmonium formation time is smaller than the time spent by the $c\overline c$ in the nucleus, defined as $\tau_{\rm c}=\langle L\rangle/(\beta_{\rm z}\gamma)$~\cite{McG13}, where $\langle L\rangle$ is the average length of nuclear matter crossed by the pair and $\beta_{\rm z}$ is the velocity of the $c\overline c$ along the beam direction in the nucleus rest frame. 
Thus, at forward-$y$, the value of $\tau_{\rm c}$ is about 10$^{-4}$ fm/$c$, while at backward-$y$ the corresponding quantity 
is $\tau_{\rm c}$$\sim$7$\cdot10^{-2}$ fm/$c$. 
Estimates for $\tau_{\rm f}$ range between 0.05 and 0.15 fm/$c$~\cite{Arl00,McG13}. Under these assumptions, no break-up effects can be expected at forward-$y$, and even at backward-$y$ the similar $\tau_{\rm c}$ and $\tau_{\rm f}$ can hardly accomodate the large discrepancy observed.
\begin{figure}[htbp]
\begin{center}
\includegraphics*[width=0.45\textwidth]{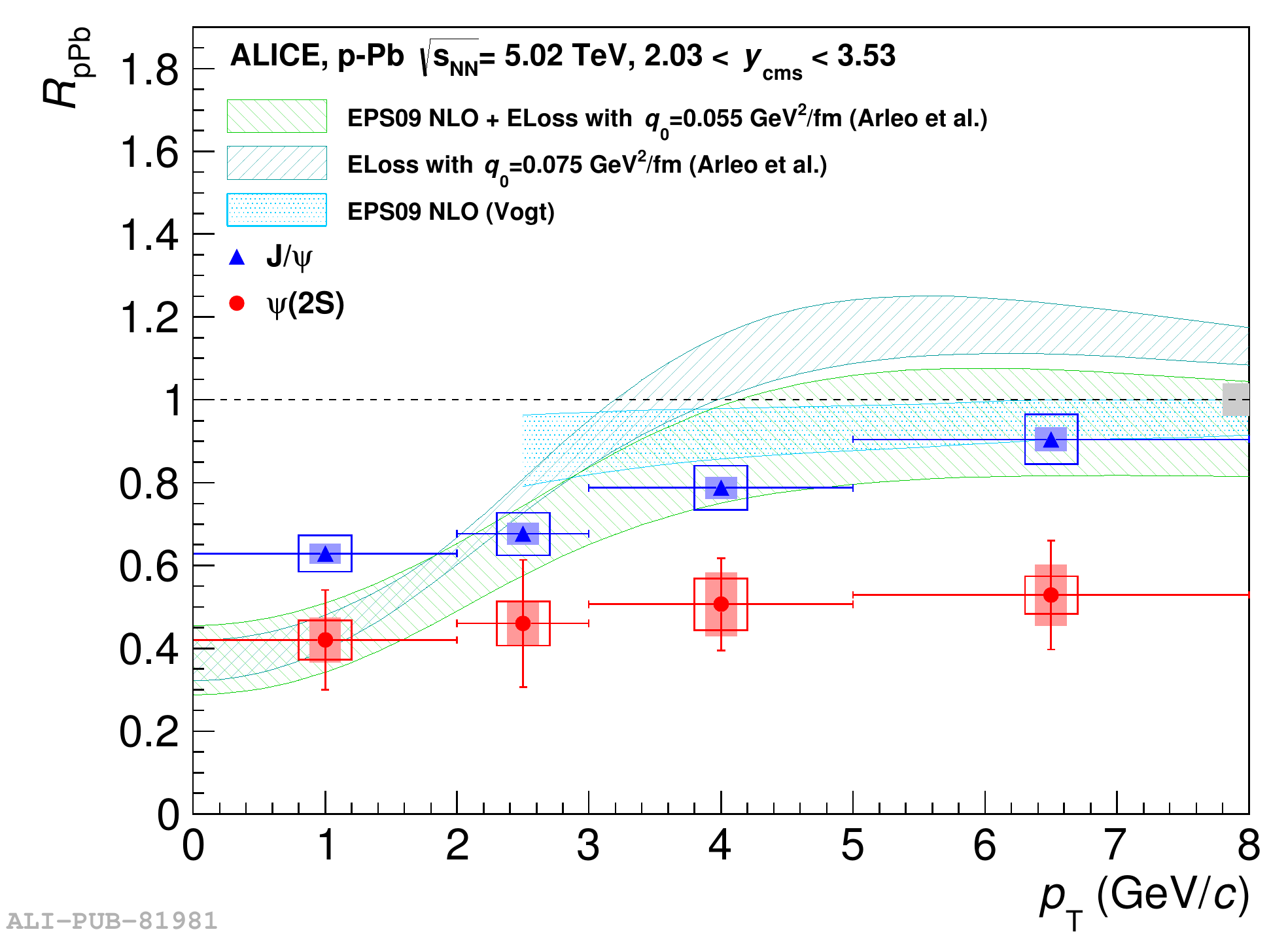}
\includegraphics*[width=0.45\textwidth]{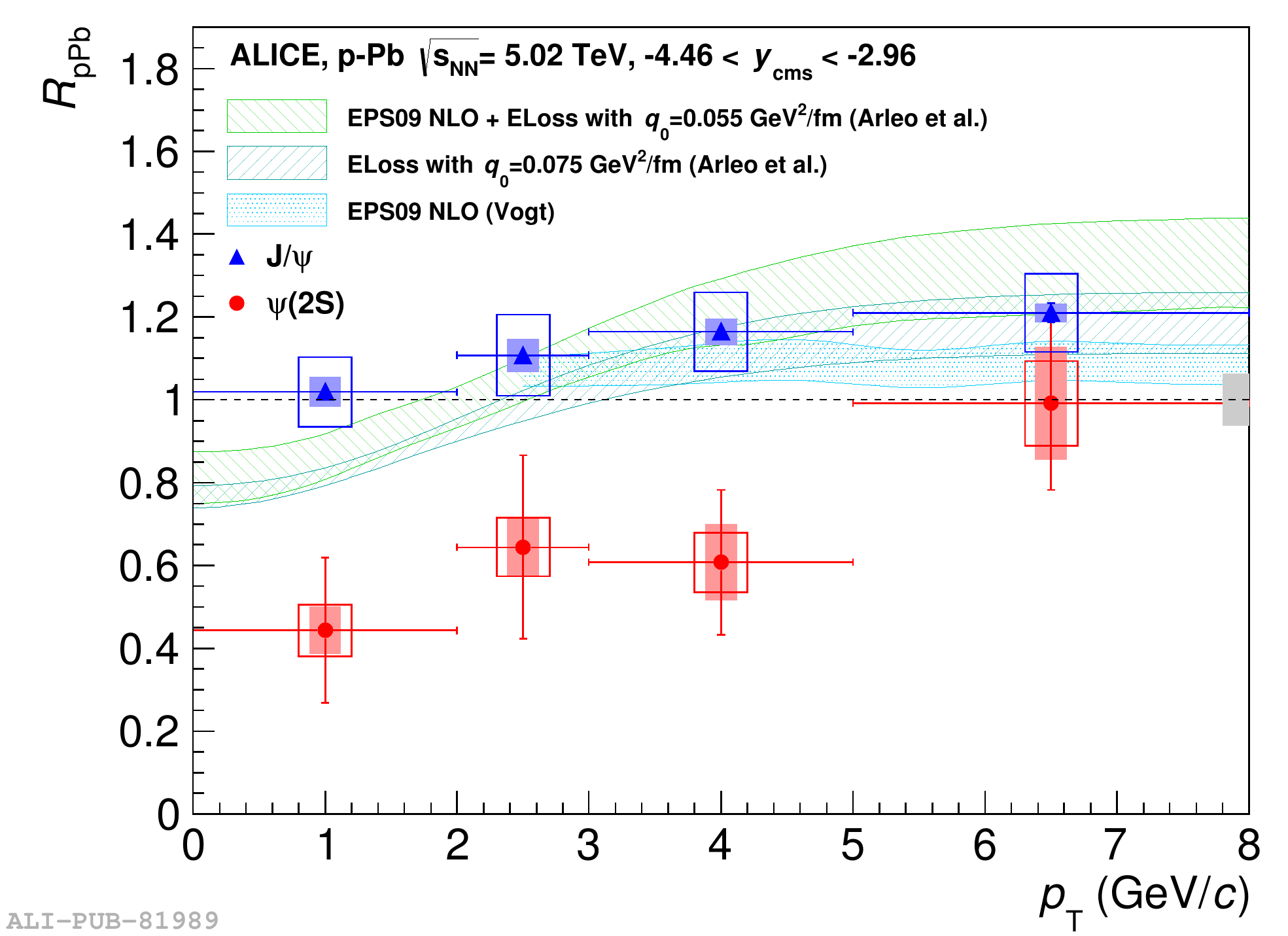}
\caption{
\psip\ and \jpsi\ \RpPb\ \pt\ dependence compared to theoretical calculations in the forward (left) and backward (right) $y$ regions. The uncertainty definition is the same as in Fig.~\ref{fig:fig1}(right).}
\label{fig:fig2}
\end{center}
\end{figure}

The \psip\ production has also been investigated as a function of the transverse momentum (\pt). The \RpPb\ computed at backward and forward $y$ are shown versus \pt\ in Fig.~\ref{fig:fig2}. 
The \psip\ \RpPb\ shows at both rapidities a strong suppression, with a slightly more evident \pt\ dependence at backward-$y$.
The \psip\ is more suppressed than the \jpsi, as already observed for the \pt-integrated 
result. Theoretical calculations, that fairly describe the \jpsi\ behaviour, overestimate the \psip\ \RpPb\ pattern.
As the \pt\ increases, the $c\overline c$ pair crossing time decreases and, in particular for backward production, $\tau_{\rm c}$ varies by a factor 2 from $\sim$0.07 (at $p_{\rm T}=0$) to $\sim$0.03 fm/$c$ (at $p_{\rm T}=8$ GeV/$c$). The role of the break-up is therefore expected to be more important at low \pt, but, even if a hint for an increasing trend can be seen in Fig.~\ref{fig:fig2} (right), no firm conclusion can be drawn from the data given the  experimental uncertainties. 

\begin{figure}[htbp]
\begin{center}
\includegraphics*[width=0.45\textwidth]{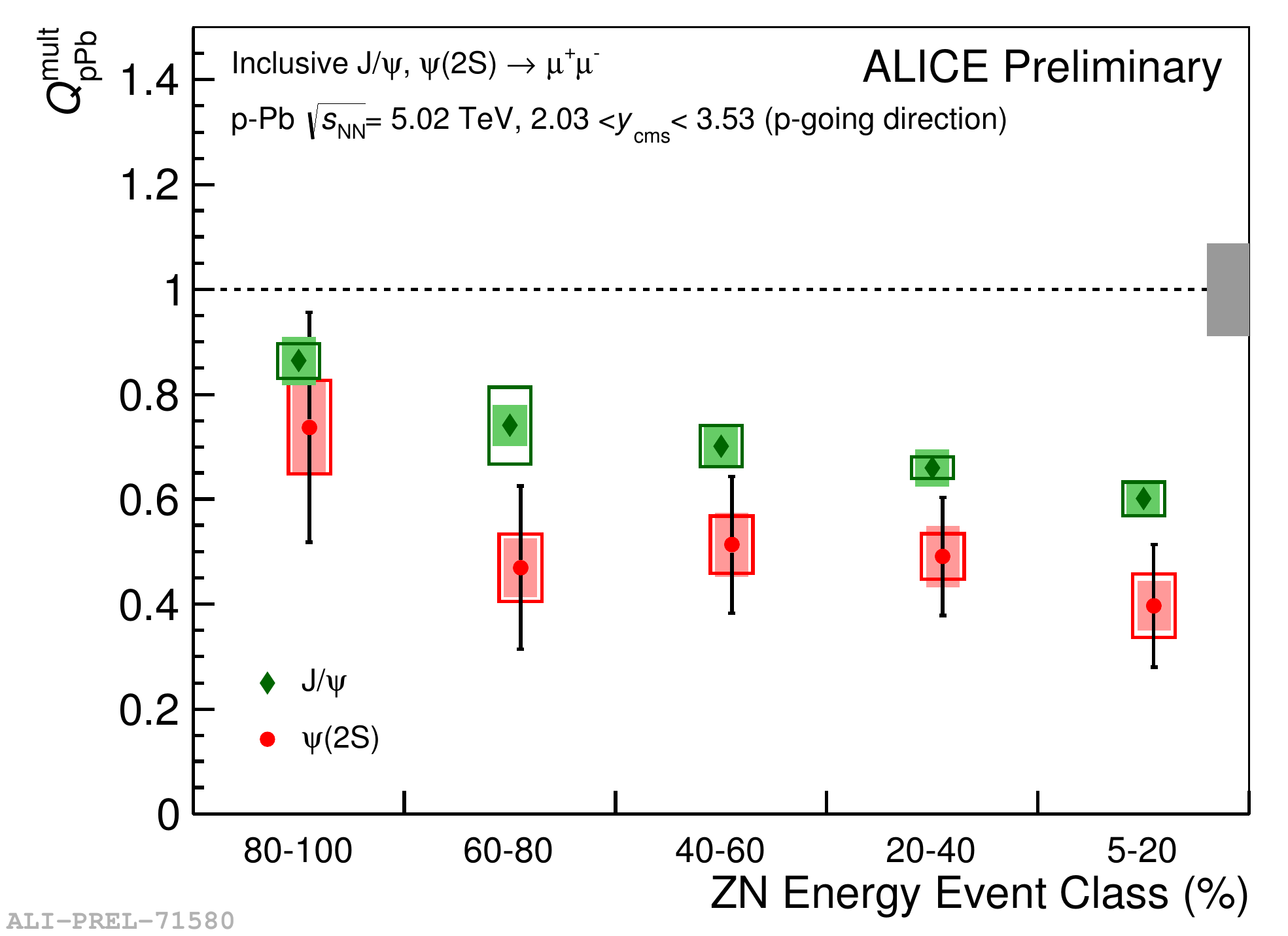}
\includegraphics*[width=0.45\textwidth]{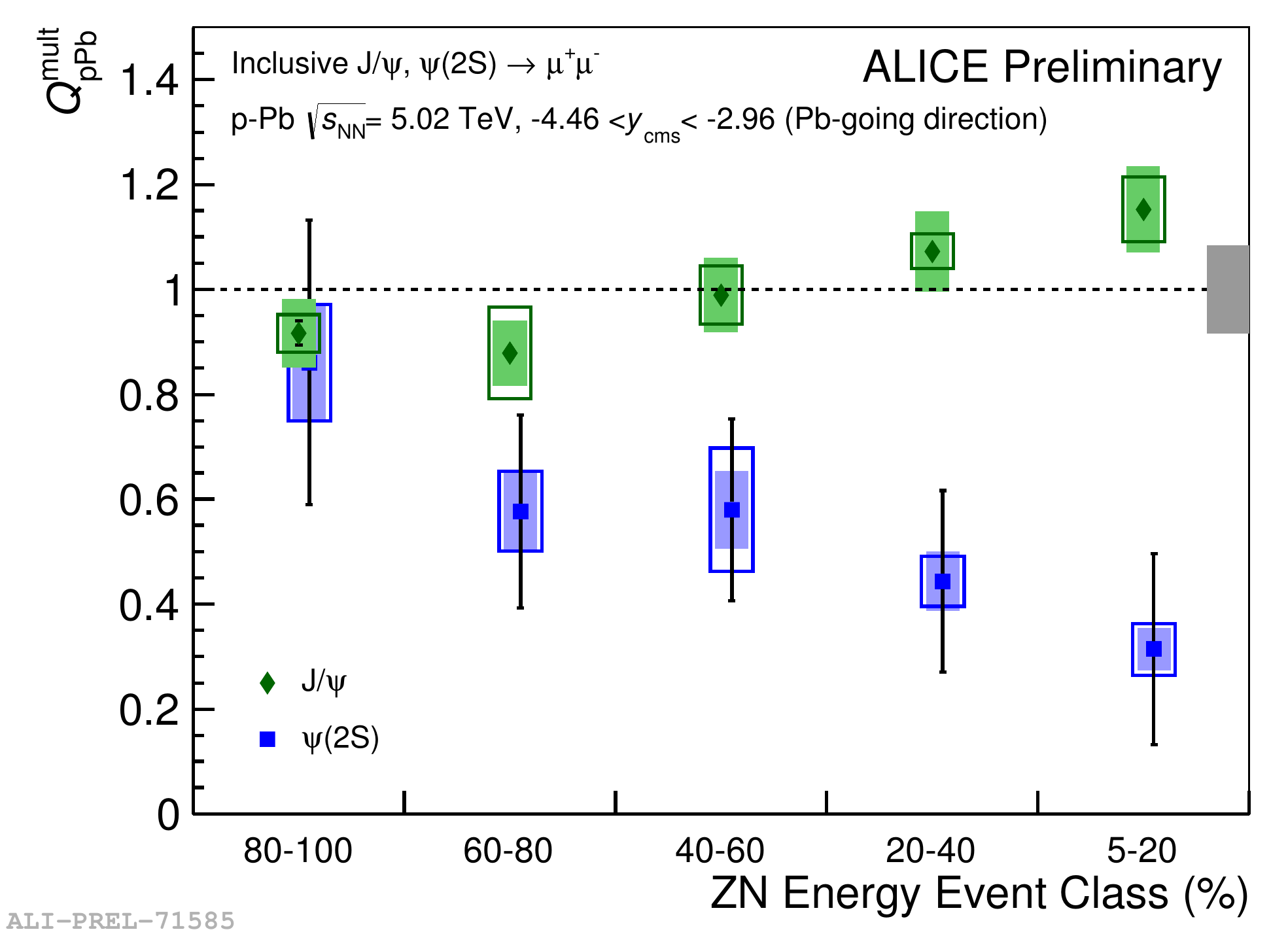}
\caption{
\jpsi\ and \psip\ \QpPb\ versus event activity in \mbox{p-Pb} at forward (left) and backward (right) $y$. The uncertainties definition is the same as in Fig.~\ref{fig:fig1} (right). The bin 0-5\% is not shown since it might by largely contaminated by pile-up.}
\label{fig:fig3}
\end{center}
\end{figure}

Finally, the \psip\ production is studied versus the event activity, as shown in Fig.~\ref{fig:fig3}. 
The event activity is determined by sampling the energy released in the neutron ZDC (ZN) and  
the link with the geometry of the collisions is established assuming, for example, that the mid-$y$ particle multiplicity scales with the number of participants nucleons~\cite{Alb04}.
To underline the fact that the centrality determination in \mbox{p-Pb} collisions can be biased by the choice of the estimator, the nuclear modification factor is, in this case, named \QpPb~\cite{Alb04}.
The \psip\ \QpPb\ shows a strong suppression, which increases with increasing event activity, and is  
rather similar in both the forward and the backward $y$ regions.
The \jpsi\ \QpPb\ shows a similar decreasing trend at forward-$y$ as a function of the event activity. 
On the contrary, the \jpsi\ and \psip\ \QpPb\ patterns, observed at backward-$y$, are rather different, with the \psip\ significantly more suppressed for large event activity classes, again pointing to additional final state effects suppressing the most weakly bound \psip. 
\section{Conclusions}
We have presented the inclusive \psip\ production in \mbox{p-Pb} collisions at \sqrts. The double ratio $[\sigma_{\psi(\rm 2S)}/\sigma_{\rm J/\psi}]_{\rm pPb}/[\sigma_{\psi(\rm 2S)}/\sigma_{\rm J/\psi}]_{\rm pp}$ and the 
\psip\ and \jpsi\ \RpPb\ (\QpPb), obtained at forward and backward $y$, have been studied as a function of \pt\ (event activity). 
Both quantities indicates that the \psip\ is significantly more suppressed than the \jpsi, with a feeble \pt\ dependence, but with a visible decrease towards 
large event activity classes. 
Initial state effects alone cannot account for the observed difference between the \jpsi\ and the \psip\ behaviour and final state effects 
as the break-up by interaction with cold nuclear matter seems unlikely, given the short time spent by the $c\overline c$ pair in the medium, in particular at forward rapidity.  Other final state effects as 
$c\overline c$ interaction with the final state hadronic medium created in \mbox{p-Pb} collisions 
should probably be considered to explain the unexpected observation.







\end{document}